\documentclass{article}

\usepackage{PRIMEarxiv}

\usepackage[utf8]{inputenc} 
\usepackage[T1]{fontenc}    
\usepackage[hidelinks]{hyperref}
\usepackage{url}            
\usepackage[numbers,sort&compress, super]{natbib}
\usepackage{natmove}
\usepackage{booktabs}       
\usepackage{amsfonts}       
\usepackage{nicefrac}       
\usepackage{microtype}      
\usepackage{lipsum}
\usepackage{fancyhdr}       
\usepackage{graphicx}       
\graphicspath{{media/}}     

\title{Bloom filters for molecules}

\author{
 Jorge Medina \\
  Department of Chemical Engineering\\
  University of Rochester \\
  Rochester, New York, USA\\
  jmedina9@ur.rochester.edu \\
   \And
  Andrew D. White \thanks{Corresponding Author} \\
    Department of Chemical Engineering\\
   University of Rochester\\
  Rochester, New York, USA\\
  andrew.white@rochester.edu \\
}

\begin{document}

\maketitle

\begin{abstract}

Ultra-large chemical libraries are reaching 10s to 100s of billions of molecules. A challenge for these libraries is to efficiently check if a proposed molecule is present.  Here we propose and study Bloom filters for testing if a molecule is present in a set using either string or fingerprint representations. Bloom filters are small enough to hold billions of molecules in just a few GB of memory and check membership in sub milliseconds. We found string representations can have a false positive rate below 1\% and require significantly less storage than using fingerprints. Canonical SMILES with Bloom filters with the simple FNV hashing function provide fast and accurate membership tests with small memory requirements. We provide a general implementation and specific filters for detecting if a molecule is purchasable, patented, or a natural product according to existing databases at \url{https://github.com/whitead/molbloom}.
\end{abstract}


\section{Introduction}


With the growing scale of molecular screening, which now involves searching through billions of chemical structures, the processing times for querying extensive compound datasets have significantly increased \cite{rester_virtuality_2008,irwin2020zinc20}. To address this, Bloom filters can compact any database just for membership verification.

The Bloom filter, a space-efficient and probabilistic data structure, was designed to ascertain whether an element belongs to a specific set. First proposed by Bloom \cite{Bloom}, this data structure has demonstrated exceptional value for large datasets, where traditional set membership testing methods would be excessively time-consuming. At its core, the Bloom filter utilizes a fixed-size ($m$) bit array to represent $n$ elements, employing $k$ hash functions to map each element to $k$ positions within the array \cite{Bloom,BloomFiltersIEEE,broder_network_2004}. This allows Bloom filters to conduct set membership tests with low false positive rates while utilizing less time and space compared to traditional data retrieval techniques.

Originally applied in dictionaries and spell checkers \cite{Bloom, SpellingList}, Bloom filters allowed for the quick identification of words within a given vocabulary, where the only significant drawback was with fake positives when misspelled words were labeled as being correct. Over time, the scope of their applications broadened to encompass web searches such as Google Chrome's former implementation of a Bloom filter to detect malicious URLs \cite{chromefilter}, among other use cases \cite{dasgupta_neural_2018,talbot_what_2015,goodwin_bitfunnel_2017}. As underscored by the Bloom Filter principle \cite{broder_network_2004}, "Wherever a list or set is used, and space is at a premium, consider using a Bloom filter if the effect of false positives can be mitigated."


 Traditionally, molecules have been represented using structure-based fingerprints \cite{fingerprintsoverview}. In this study, we built different bloom filters using the Coconut database \cite{sorokina_coconut_2021} to compare the effectiveness of structure-based hashing with string hashes in the Bloom filter; we demonstrate that string hashing consistently outperforms its counterpart. To provide further context, Table \ref{tab:table} presents well-known chemistry databases, their approximate number of compounds, storage size required for text (SMILES) representation, and a comparison with a Bloom filter designed to store an equivalent number of molecules.


This study explores different ways of using bloom filters for molecules. It is important to note that, while not a focus of our investigation, alternative data structures address some limitations inherent to Bloom filters. Notably, Cuckoo filters \cite{cuckoo} allow for dynamic item insertion and deletion, a feature lacking in traditional Bloom filters. In addition to Cuckoo filters, other alternatives such as Quotient filters \cite{quotient} and Count-Min sketches \cite{count-min} can be found in the literature. 
\begin{table}
 
  \centering
  \begin{tabular}{llcc}
    \toprule
    Name     & Size (\# of compounds) & Size\ (GB)$^a$ &  Bloom  filter  size needed (GB)$^{b}$  \\
    \midrule
    ZINC\cite{irwin2020zinc20}        & $>2$ billion  & >100   &   >2.56730  \\
    ChemBL\cite{gaulton_chembl_2012}      & $2,354,965$    & 0.117  &   0.003023    \\
    Coconut\cite{sorokina_coconut_2021}     & $407,270$      & 0.0204 &   0.000522 \\
    BindingDB\cite{liu_bindingdb_2007}   & $566,000$      & 0.0283 &   0.000727  \\
    PubChem\cite{kim_pubchem_2023}     & $113,993,087$  & 5.700  &   0.146344  \\
    SureChemBL\cite{papadatos_surechembl_2016}  & $22,843,364$   & 1.1422 &   0.02932    \\
    Available Chemical Directory & $>3,2$ million & > 0.16  & >0.004108\\
    ChemNavigator & $10,000,000$  & 0.5   & 0.012837 \\
    ChemBridge    & $1,3$ million &  0.065 & 0.001668  \\
    ChemSpider\cite{pence_chemspider_2010}    & $>115,000,000$ & >5.75  & >0.14763 \\
    \bottomrule\\
  \end{tabular}
  \caption{Examples of Chemical Compounds Databases. 
  $^a$ Estimated text file sizes for SMILES based on a 95M sample; total sizes inferred by linear scaling to avoid loading entire datasets.
  $^b$ Estimated Bloom filter size needed for a fixed false positive rate of 0.005.}
  \label{tab:table}
\end{table}

A Bloom filter is initialized with an m-length bit vector, with all positions set to zero, and employs k independent hashing functions. These hashing functions generate k values ranging from 0 to m$-$1, which correspond to the positions in the bit vector where a "1" will be assigned. The hashing functions must exhibit the following characteristics:\cite{hashfuncs}

\begin{enumerate}
    \item Quick computation,
    \item An avalanche effect, where minor input changes result in substantial and unpredictable output alterations,
    \item The generation of integers between 0 and m$-$1.
\end{enumerate}

\begin{figure}[htp]
    \centering
    \includegraphics[width=7cm]{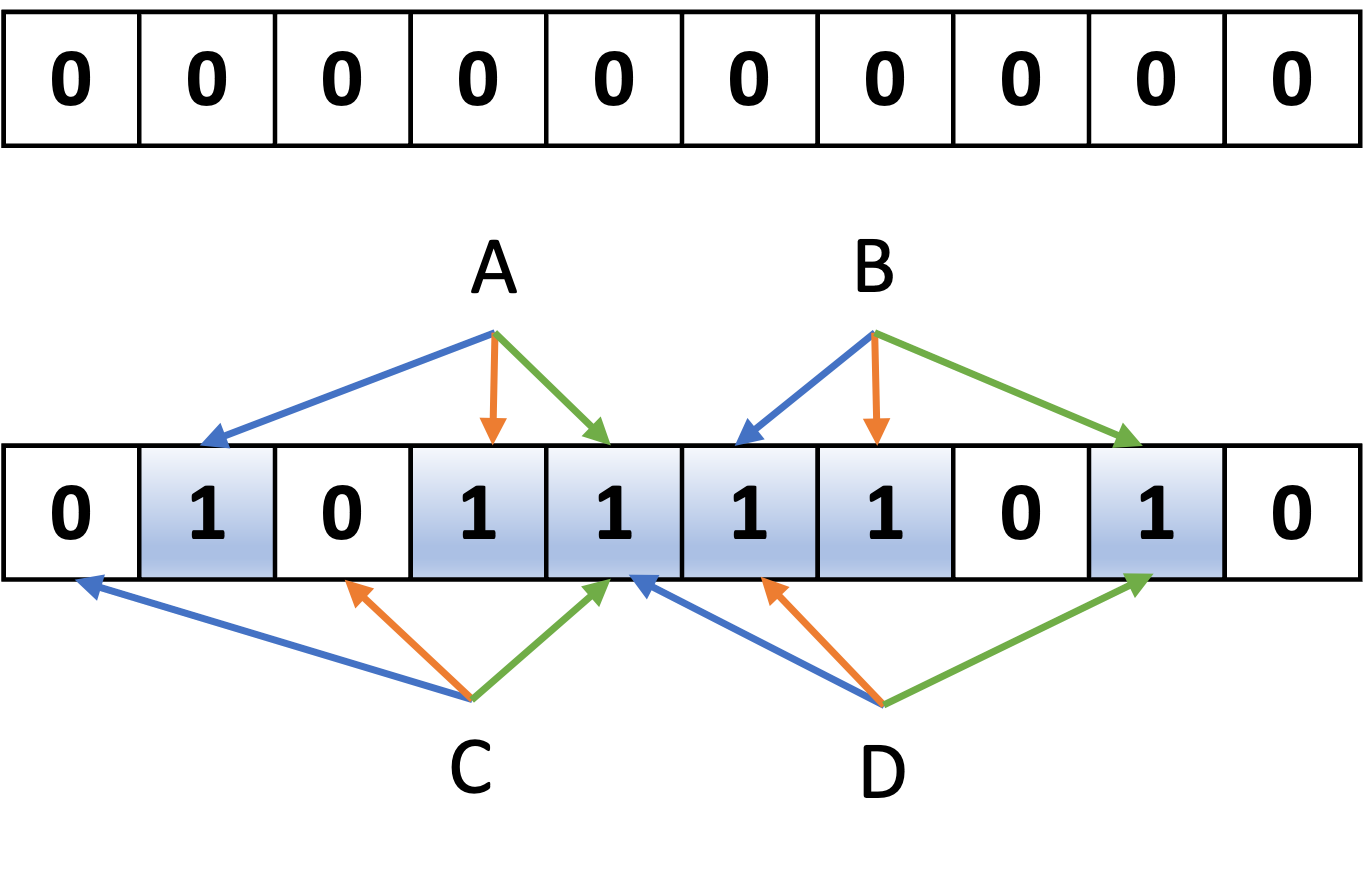}
    \caption{Scheme of Bloom Filters. In this generic Bloom filter example, we start with an empty bit array of zeros and four elements: A, B, C, and D. The first two elements (A and B) are added to the filter, while the latter two (C and D) are queried. The process utilizes three distinct hashing functions, represented by colored arrows. To verify if elements C and D have been previously added to the filter, they are checked using these hashing functions. For element C, one of the hashing functions points to a zero bit, indicating that the element has not been added to the filter. However, all three hashing functions for element D point to bits already set to one, resulting in a false positive.}
    \label{fig:BloomFilterexample}
\end{figure}

\begin{figure}[htp]
    \centering
    \includegraphics[width=16cm]{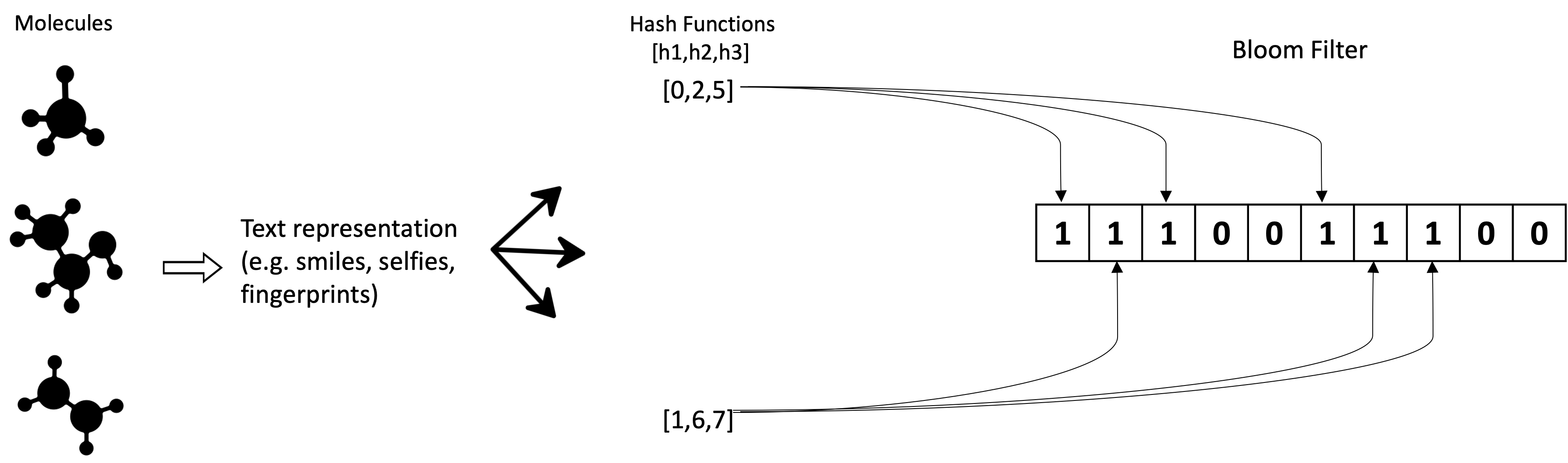}
    \caption{Illustrative Example for Bloom Filter with Molecules. When populating a Bloom filter, a set of molecules is initially stored in either a text format such as SMILES or SELFIES, or as fingerprints. Subsequently, distinct hashing functions generate indexes (three in this example) for each element to fill the filter.}
    \label{fig:galaxy}
\end{figure}

Bloom filters enable the addition of new members but do not support individual removals. The filter can be queried to determine if a particular element has been added previously. However, this simplicity comes with certain drawbacks, such as the potential for two or more elements to be hashed to the same position in the Bloom filter (i.e., collisions). As a result, removing an element (by changing its positions from one to zero) could inadvertently affect other members with overlapping positions. This issue underscores the importance of randomness in hashing functions, often referred to as the avalanche effect.  Figures \ref{fig:BloomFilterexample} and \ref{fig:galaxy} illustrate the workings of a Bloom filter and the storage of molecules within such filters.

For more comprehensive information on Bloom filter functionality, theoretical limits, and optimal implementation, refer to the existing literature. \cite{wikipedia-contributors-2023,BloomFiltersIEEE,Bloom,broder_network_2004}

Double hashing is employed to minimize the probability of collisions in the indexing of new members \cite{double_hashing}. Two distinct "universal hashes," $h_\alpha$ and $h_\beta$, are utilized to obtain n individual indices:

 $$h_i(A) = (h_\alpha(A) + i * h_\beta(A)) \ mod \ |m| $$
Here, `A' represents an element being hashed, $h_i$ refers to the k hash functions generated per element (as illustrated in Figures \ref{fig:BloomFilterexample} and \ref{fig:galaxy}), |m| denotes the fixed size of the filter, and "mod" signifies the remainder of the division. Restrictions that reduce collision probability are \cite{double_hashing}:

\begin{itemize}
    \item $h_\beta \neq 0$,
    \item $h_\beta(A)$ should not be divisible-like by the size of the filter. 
\end{itemize}

Using the described method, we generate an arbitrary number of hashing functions.

\section{Methodology}
The Python package MolBloom developed for this work \cite{White_molbloom_quick_assessment_2022} is an open-source package designed for molecules, featuring a built-in filter with ZINC-in-stock molecules. The package permits the creation of custom filters of varying sizes, which were adjusted in increments of one order of magnitude. Tests were conducted using the Coconut dataset \cite{sorokina_coconut_2021} (approximately 400,000 molecules).

For comparative purposes, molecular fingerprints were employed to populate a Bloom filter and measure the false positive rate for increasing bit-array sizes. The hashing functions used in this study include Fowler-Noll-Voll (FNV) \cite{eastlake-fnv-19}, as well as message digest 4 and 5 algorithms (md4 and md5) \cite{md4,md5} for string hashing. For chemical structure fingerprints, six combination between MACCS \cite{MAACS}, Morgan \cite{Morgan}, Atom-pair \cite{atompairfp}, and RDKit Fingerprints were utilized. This was done to investigate how traditional ways to hash molecules would act in this setting. FNV is a hash function designed for rapid, non-cryptographic hashing of data, leveraging prime numbers and bitwise operations to generate hash values that identify unique data elements. The FNV algorithm offers variants of different bit sizes and prime numbers, such as FNV-1 and FNV-1a. MD4 and MD5 are well-established hashing functions within the computer science community \cite{Bosselaers2005}.

To assess false positive rates in each filter with different sizes, a fifty-fifty split was performed. The first half was added to empty filters, followed by membership testing in the second half. Any molecules from the second half classified as part of the set were counted as false positives.

An evaluation was conducted to compare the speed of Bloom filters and traditional methods in searching for elements within a dataset (using the dataset's native API). 

\section{Results and Discussion}

\begin{figure}[h]
    \centering
        \centering
        \includegraphics[width=1\linewidth]{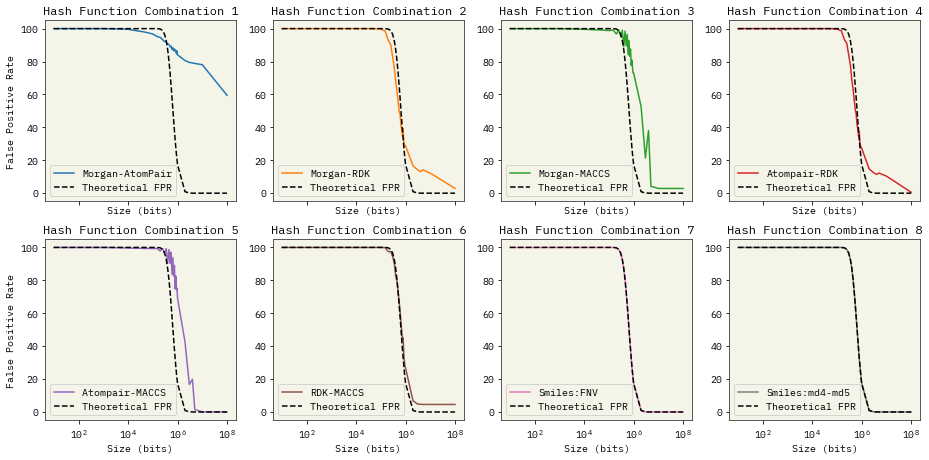}%
    \caption{False Positive Rate vs filter size for different hash methods. Although all hashing variations follow similar trends, both string hashes are identical. ``Noisy'' peaks can be seen, which result from hashing functions being divisible-like by the size of the bloom filter}
    \label{fig:results}
\end{figure}

All six possible fingerprint combinations across eight distinct orders of magnitude for the Bloom filter and string hash implementations were examined. Figure \ref{fig:results} provide a comprehensive summary of the results.

As illustrated in Figure \ref{fig:results}, two key observations can be made. First, as anticipated, the false positive rate of Bloom filters approaches zero as the ratio between the filter size and dataset size increases. Second, the hashing of string SMILES representation outperforms most chemical structure fingerprints by over an order of magnitude in terms of false positive rate (combinations 7 \& 8). Only the Morgan-MACCS and Atompair-MACCS fingerprint (combinations 3 \& 5) hashing achieve false positive rates comparable to strings while requiring half an order of magnitude more bits of space.

Message Digest and FNV hashing (7 \& 8) of strings yielded nearly identical and seemingly smooth curves, suggesting a well-randomized hashing of the elements. In contrast, other methods exhibit a "noisy" pattern, which serves as evidence of inadequate randomization. By design, these alternative methods are not highly randomized, as similar molecules tend to have comparable chemical fingerprints. This characteristic is the basis for their use in numerous optimization methods, as they can measure the distance between molecules. Consequently, their performance is suboptimal, as similar molecules have a higher likelihood of collisions within the Bloom filter.

\begin{figure}[h!]
    \centering
        \centering
        \includegraphics[width=0.55\linewidth]{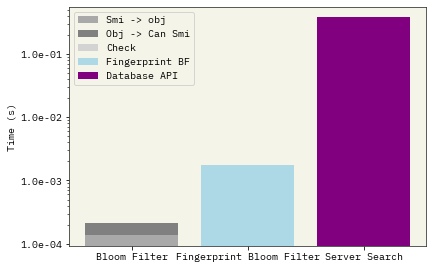}
    \caption{Comparison of time usages using bloom filters and direct search in a commercial database. For Bloom filters, the analysis was done by separating the steps necessary: 1) transform the smiles into a RDkit mol object, 2) canonicalize the string and 3) check the membership. The time needed for each can be seen in the first column. }
    \label{fig:timeresults}
\end{figure}

In terms of the time required for Bloom filters to verify whether a molecule is part of a set or not, Figure \ref{fig:timeresults} clearly illustrates that Bloom filters demand up to three orders of magnitude less time compared to the native API. Even the "slower" Python implementation using RDKit for fingerprints necessitates two orders of magnitude less time for membership checks.

\section{Conclusion}


We demonstrate that string hashing (FNV and MD4-5) for Bloom filters outperform and approximate the theoretical limit of these structures, confirming that strings are sufficient for molecule storage. Even taking into account the time spent on canonicalizing SMILES, Bloom filter retrieval is still more than two orders of magnitudes faster than using an internet search. We also show that FNV, despite its simplicity and speed, is as effective as MD5. Employing other string representations, such as InChI and SELFIES, is expected to yield similar results. Potential applications for the Bloom filter are to quickly determine if a molecule is purchasable in ZINC\cite{irwin2020zinc20}, patented according to SureChembl\cite{papadatos2016surechembl}, or a natural product\cite{sorokina_coconut_2021}. 

\section{Data and Code availability}
Molbloom package is an open source project and the code implementation in python for the experiments can be found at: \url{https://github.com/whitead/molbloom} and \url{https://github.com/Jgmedina95/molbloom-paper}.

\section*{Acknowledgments}
We thank the Center for Integrated Research Computing (CIRC) at the University of Rochester for providing computational resources and technical support.

\bibliographystyle{unsrt}
\bibliography{references}

\begin{thebibliography}{10}

\bibitem{rester_virtuality_2008}
Ulrich Rester.
\newblock From virtuality to reality - {Virtual} screening in lead discovery
  and lead optimization: a medicinal chemistry perspective.
\newblock {\em Current Opinion in Drug Discovery \& Development},
  11(4):559--568, July 2008.

\bibitem{irwin2020zinc20}
John~J Irwin, Khanh~G Tang, Jennifer Young, Chinzorig Dandarchuluun, Benjamin~R
  Wong, Munkhzul Khurelbaatar, Yurii~S Moroz, John Mayfield, and Roger~A Sayle.
\newblock Zinc20—a free ultralarge-scale chemical database for ligand
  discovery.
\newblock {\em Journal of chemical information and modeling},
  60(12):6065--6073.

\bibitem{Bloom}
Burton~H. Bloom.
\newblock Space/time trade-offs in hash coding with allowable errors.
\newblock {\em Commun. ACM}, 13(7):422–426, jul 1970.

\bibitem{BloomFiltersIEEE}
Sasu Tarkoma, Christian~Esteve Rothenberg, and Eemil Lagerspetz.
\newblock Theory and practice of bloom filters for distributed systems.
\newblock {\em IEEE Communications Surveys \& Tutorials}, 14(1):131--155, 2012.

\bibitem{broder_network_2004}
Andrei Broder and Michael Mitzenmacher.
\newblock Network {Applications} of {Bloom} {Filters}: {A} {Survey}.
\newblock {\em Internet Mathematics}, 1(4):485--509, January 2004.

\bibitem{SpellingList}
M.~McIlroy.
\newblock Development of a spelling list.
\newblock {\em IEEE Transactions on Communications}, 30(1):91--99, 1982.

\bibitem{chromefilter}
Alex Yakunin.
\newblock Nice bloom filter application, 2010.

\bibitem{dasgupta_neural_2018}
Sanjoy Dasgupta, Timothy~C. Sheehan, Charles~F. Stevens, and Saket Navlakha.
\newblock A neural data structure for novelty detection.
\newblock {\em Proceedings of the National Academy of Sciences},
  115(51):13093--13098, December 2018.

\bibitem{talbot_what_2015}
Jamie Talbot.
\newblock What are {Bloom} filters?, July 2015.

\bibitem{goodwin_bitfunnel_2017}
Bob Goodwin, Michael Hopcroft, Dan Luu, Alex Clemmer, Mihaela Curmei, Sameh
  Elnikety, and Yuxiong He.
\newblock {BitFunnel}: {Revisiting} {Signatures} for {Search}.
\newblock In {\em Proceedings of the 40th {International} {ACM} {SIGIR}
  {Conference} on {Research} and {Development} in {Information} {Retrieval}},
  pages 605--614, Shinjuku Tokyo Japan, August 2017. ACM.

\bibitem{fingerprintsoverview}
Ingo Muegge and Prasenjit Mukherjee.
\newblock An overview of molecular fingerprint similarity search in virtual
  screening.
\newblock {\em Expert Opinion on Drug Discovery}, 11(2):137--148, February
  2016.

\bibitem{sorokina_coconut_2021}
Maria Sorokina, Peter Merseburger, Kohulan Rajan, Mehmet~Aziz Yirik, and
  Christoph Steinbeck.
\newblock {COCONUT} online: {Collection} of {Open} {Natural} {Products}
  database.
\newblock {\em Journal of Cheminformatics}, 13(1):2, January 2021.

\bibitem{cuckoo}
Bin Fan, Dave~G. Andersen, Michael Kaminsky, and Michael~D. Mitzenmacher.
\newblock Cuckoo filter: Practically better than bloom.
\newblock In {\em Proceedings of the 10th ACM International on Conference on
  Emerging Networking Experiments and Technologies}, CoNEXT '14, page 75–88,
  New York, NY, USA, 2014. Association for Computing Machinery.

\bibitem{quotient}
Michael~A Bender, Martin Farach-Colton, Rob Johnson, Bradley~C Kuszmaul, Dzejla
  Medjedovic, Pablo Montes, Pradeep Shetty, Richard~P Spillane, and Erez Zadok.
\newblock Don't thrash: how to cache your hash on flash.
\newblock In {\em 3rd Workshop on Hot Topics in Storage and File Systems
  (HotStorage 11)}, 2011.

\bibitem{count-min}
Graham Cormode.
\newblock Count-min sketch., 2009.

\bibitem{gaulton_chembl_2012}
A.~Gaulton, L.~J. Bellis, A.~P. Bento, J.~Chambers, M.~Davies, A.~Hersey,
  Y.~Light, S.~McGlinchey, D.~Michalovich, B.~Al-Lazikani, and J.~P.
  Overington.
\newblock {ChEMBL}: a large-scale bioactivity database for drug discovery.
\newblock {\em Nucleic Acids Research}, 40(D1):D1100--D1107, January 2012.

\bibitem{liu_bindingdb_2007}
T.~Liu, Y.~Lin, X.~Wen, R.~N. Jorissen, and M.~K. Gilson.
\newblock {BindingDB}: a web-accessible database of experimentally determined
  protein-ligand binding affinities.
\newblock {\em Nucleic Acids Research}, 35(Database):D198--D201, January 2007.

\bibitem{kim_pubchem_2023}
Sunghwan Kim, Jie Chen, Tiejun Cheng, Asta Gindulyte, Jia He, Siqian He,
  Qingliang Li, Benjamin~A Shoemaker, Paul~A Thiessen, Bo~Yu, Leonid Zaslavsky,
  Jian Zhang, and Evan~E Bolton.
\newblock {PubChem} 2023 update.
\newblock {\em Nucleic Acids Research}, 51(D1):D1373--D1380, January 2023.

\bibitem{papadatos_surechembl_2016}
George Papadatos, Mark Davies, Nathan Dedman, Jon Chambers, Anna Gaulton, James
  Siddle, Richard Koks, Sean~A. Irvine, Joe Pettersson, Nicko Goncharoff, Anne
  Hersey, and John~P. Overington.
\newblock {SureChEMBL}: a large-scale, chemically annotated patent document
  database.
\newblock {\em Nucleic Acids Research}, 44(D1):D1220--D1228, January 2016.

\bibitem{pence_chemspider_2010}
Harry~E. Pence and Antony Williams.
\newblock {ChemSpider}: {An} {Online} {Chemical} {Information} {Resource}.
\newblock {\em Journal of Chemical Education}, 87(11):1123--1124, November
  2010.

\bibitem{hashfuncs}
Tom {St Denis} and Simon Johnson.
\newblock Chapter 5 - hash functions.
\newblock In Tom {St Denis} and Simon Johnson, editors, {\em Cryptography for
  Developers}, pages 203--250. Syngress, Burlington, 2007.

\bibitem{wikipedia-contributors-2023}
Wikipedia contributors.
\newblock {Bloom filter}, 2 2023.

\bibitem{double_hashing}
Peter C~Dillinger <peterd@cc.gatech.edu> Panagiotis
  Manolios~<manolios@cc.gatech.edu>.
\newblock Bloom filters in probabilistic verification.
\newblock {\em International Conference on Formal Methods in Computer-Aided
  Design}, 2004.

\bibitem{White_molbloom_quick_assessment_2022}
Andrew~D White.
\newblock {molbloom: quick assessment of compound purchasability with bloom
  filters}, 12 2022.

\bibitem{eastlake-fnv-19}
Glenn Fowler, Landon~Curt Noll, Kiem-Phong Vo, Donald E.~Eastlake 3rd, and Tony
  Hansen.
\newblock {The FNV Non-Cryptographic Hash Algorithm}.
\newblock Internet-Draft draft-eastlake-fnv-19, Internet Engineering Task
  Force, January 2023.
\newblock Work in Progress.

\bibitem{md4}
Ronald~L. Rivest.
\newblock {The MD4 Message-Digest Algorithm}.
\newblock RFC 1320, April 1992.

\bibitem{md5}
Ronald~L. Rivest.
\newblock {The MD5 Message-Digest Algorithm}.
\newblock RFC 1321, April 1992.

\bibitem{MAACS}
Joseph~L. Durant, Burton~A. Leland, Douglas~R. Henry, and James~G. Nourse.
\newblock Reoptimization of mdl keys for use in drug discovery.
\newblock {\em Journal of Chemical Information and Computer Sciences},
  42(6):1273--1280, 2002.
\newblock PMID: 12444722.

\bibitem{Morgan}
H.~L. Morgan.
\newblock The generation of a unique machine description for chemical
  structures-a technique developed at chemical abstracts service.
\newblock {\em Journal of Chemical Documentation}, 5(2):107--113, 1965.

\bibitem{atompairfp}
Alice Capecchi, Daniel Probst, and Jean-Louis Reymond.
\newblock One molecular fingerprint to rule them all: drugs, biomolecules, and
  the metabolome.
\newblock {\em Journal of Cheminformatics}, 12(1):43, December 2020.

\bibitem{Bosselaers2005}
Antoon Bosselaers.
\newblock {\em Md4-Md5}, pages 378--379.
\newblock Springer US, Boston, MA, 2005.

\bibitem{papadatos2016surechembl}
George Papadatos, Mark Davies, Nathan Dedman, Jon Chambers, Anna Gaulton, James
  Siddle, Richard Koks, Sean~A Irvine, Joe Pettersson, Nicko Goncharoff, et~al.
\newblock Surechembl: a large-scale, chemically annotated patent document
  database.
\newblock {\em Nucleic acids research}, 44(D1):D1220--D1228, 2016.

\end{thebibliography}

\end{document}